\documentclass[review]{elsarticle}

\usepackage{amsmath}
\usepackage{graphicx}
\usepackage{lineno,hyperref}
\journal{Carbon}

\begin{document}

\begin{frontmatter}

\title{Taming the magnetoresistance anomaly in graphite \footnote{This manuscript version is made available under the CC-BY-NC-ND 4.0 license http://creativecommons.org/licenses/by-nc-nd/4.0/}}

%\author{Bruno Cury Camargo}

\author[aff_1,aff_2]{Bruno Cury Camargo\corref{mycorrespondingauthor}}
\cortext[mycorrespondingauthor]{Corresponding author. Tel. +48-22-1163334}
\ead{b.c\_camargo@yahoo.com.br}
%\phone{+48-22-1163334}

\author[aff_2]{Walter Escoffier}
%\ead{walter.escoffier@lncmi.cnrs.fr}

\address[aff_1]{Institute of Physics, Polish Academy of Sciences, Aleja Lotnikow 32/46, PL-02-668 Warsaw, Poland.}
\address[aff_2]{Laboratoire National des Champs Magnetiques Intenses, CNRS-INSA-UJF-UPS, UPR3228; 143 avenue de Rangueil, F-31400 Toulouse, France.}

\begin{abstract}
At low temperatures, graphite presents a magnetoresistance anomaly which manifests as a transition to a high-resistance state (HRS) above a certain critical magnetic field $\text{B}_\text{c}$. Such HRS is currently attributed to a c-axis charge-density-wave taking place only when the lowest Landau level is populated. By controlling the charge carrier concentration of a gated sample through its charge neutrality level (CNL), we were able to experimentally modulate the HRS in graphite for the first time.  We demonstrate that the HRS is triggered both when electrons and holes are the majority carriers but is attenuated near the CNL. Taking screening into account, our results indicate that the HRS possess a strong in-plane component and can occur below the quantum limit, being at odds with the current understanding of the phenomenon. We also report the effect of sample thickness on the HRS.
\end{abstract}

%\begin{keyword}
%\texttt{elsarticle.cls}\sep \LaTeX\sep Elsevier \sep template
%\MSC[2010] 00-01\sep  99-00
%\end{keyword}

\end{frontmatter}

%\linenumbers

\section{Introduction}

Graphite is a quasi-compensated semimetal in which charge carriers possess high electronic mobility and low effective masses \cite{McClure1964}. These allow the material to reach the quantum limit at modest values of $\text{B} \approx 7$ T (the smallest magnetic field for which only the lowest energy Landau level (LL) remains populated) \cite{Stamenov2005}. At higher magnetic fields however, graphite hints at some exotic properties, such as the occurrence of the fractional quantum Hall effect, the possibility of magnetic-field-induced superconductivity and the existence of a magnetic-field-induced high resistance state (HRS) \cite{Kopelevich1999, Kopelevich2009, Iye1982}. The latter has been thoroughly investigated since its first experimental observation in the 1980's , and still sparks off debate to date \cite{Iye1982, Timp1983, Nakamura1983, Yaguchi1998, Kumar2010, Fauque2013}. It manifests as a single or multiple sharp bump(s) of the sample resistance as a function of magnetic field, usually triggered at $\text{B} > 25$ T \cite{Timp1983, Nakamura1983, Yaguchi1998, Kumar2010, Fauque2013}.

Much experimental work has been devoted to verify the origin of this state (see ref. \cite{Yaguchi2009} for a review). Despite earlier reports suggesting that the HRS is an in-plane phenomenon, current consensus is that it is triggered along the c-axis direction at the lowest Landau level in graphite \cite{Timp1983, Takada1998}. Early theoretical attempts by Yoshioka and Fukuyama invoked the surging of a charge-density-wave (CDW) transition along the sample c-axis, caused by a 3D to 1D dimensionality reduction due to the quantum limit\cite{Yoshioka1981}. In this context, the occurrence and subsequent suppression of the CDW state have been attributed to the crossing between the Fermi level and the lowest Landau spin subbands at increasing magnetic fields.

Albeit it is widely accepted that the effect is caused by electron-electron interactions, experimentalists still struggle to verify the physical mechanisms responsible for the HRS. Currently, reports support different hypothesis, which include the formation of a CDW, a spin-density wave or the opening of an excitonic gap. All expected to occur along graphite's c-axis at the lowest Landau level \cite{Kumar2010, Fauque2013, Akiba2015, Arnold2017, Takahashi1994}.  

Different approaches to understand the nature of the HRS have been attempted in the past decades, highlighting various aspects of the phenomenon. For example, early angle-resolved measurements by Timp et al. showcased the c-axis component of magnetic field as the sole responsible for the HRS, supporting the presence of a CDW in graphite. Similar experiments recently performed by Zhu et al., however, showed a suppression of the HRS with tilting angle, weighting in favour of an excitonic gap taking place \cite{Timp1983, Zhu2017}. As another example, although earlier models ascribed the HRS to a CDW triggered by a single spin subband in the lowest Landau level of graphite \cite{Yoshioka1981}, recent measurements by Fauque et al. at $\text{B} = 80$ T demonstrated a re-entrant behavior, suggesting its triggering by more than one spin subband \cite{Fauque2013}. 

Although such works provide tantalizing hints of the HRS origin, current experimental data does not allow for an indisputable implication of the bands and carriers responsible for the phenomenon. One solution for this problem would be to monitor the HRS for doped graphite samples. In doing so, one could actively depopulate different spin subbands of electrons and holes by changing the sample's chemical potential. To our best knowledge, attempts in this topic are currently carried out by performing ionic implantation in bulk graphite or testing different sample qualities. These approaches do not allow one to reliably separate the effects caused by induced disorder from the ones caused by the intended doping \cite{Yaguchi2009, Akiba2015}. This becomes critical when comparing different samples, as differences on disorder and native charge carrier concentration can affect the phenomenon under consideration, causing seemingly similar experiments to produce potentially diverging conclusions. 

In order to address this issue, in the present work, we verify the behavior of the HRS in mesoscopic graphite by controlling the material's chemical potential (i.e. the position of the Fermi level within the band structure) by gating our device through its Charge Neutrality Level (CNL). In doing so, we were able to electrostatically dope the sample without modifying other critical parameters (such as cristalinity, geometry and quality of contacts). Our results provide the first experimental observation that the properties of the HRS change non-monotonically with graphite's averaged charge carrier density. We observe that the HRS survives outside the quantum limit and is triggered by both electrons and holes, approximately symmetric to each other with respect to the CNL. By accounting for charge screening we also infer a strong two-dimensional character of the HRS, although an off-plane degree of freedom seems necessary for the phenomenon. Our results shed new light on the subject, suggesting that the HRS in graphite has a large in-plane component and might take place at Landau levels with $n>0$.

\section{Results and discussion}

\begin{figure}[h]
	\includegraphics[width=8cm]{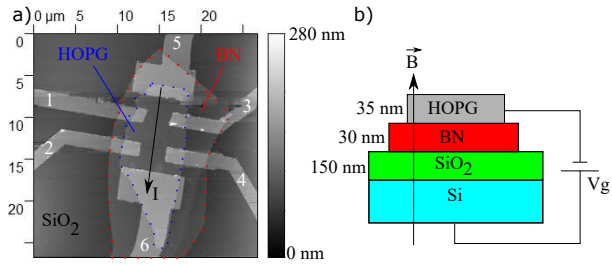}
	\caption{a) AFM image of the HOPG sample studied here. The blue (red) dots mark the boundaries of the HOPG (BN) in the device. The numbers on the figure are used to identify the contacts. The electrical current is applied between contacts $5$ and $6$. $\text{R}_\text{xx}$ is measured between contacts $1$ and $2$ and $\text{R}_\text{xy}$ between contacts $2$ and $4$. b) Cross-section schematic of the device, showing how the backgate voltage was applied and the direction of the magnetic field.}
	\label{fig_picture}
\end{figure}

The experiments shown here were carried out in mesoscopic Highly Oriented Pyrolytic Graphite (HOPG) exfoliated from a bulk crystal with mosaicity of $0.30$ \cite{GW_INC} - the Full width at Half Maxima (FWHM) extracted from x-ray rocking-curve measurements. The device had approximate in-plane dimensions of $5$ $\mu$m $\times$ $5$ $\mu$m (see figure \ref{fig_picture}) and a thickness of $35$ nm. The sample was deposited atop a $30$ nm-thick BN crystal previously placed on a N-dopped Si substrate coated with 300 nm of $\text{SiO}_2$. The sample was contacted with electron-beam lithography for longitudinal and Hall measurements. A backgate voltage in the range \mbox{$-30$ V $\leq$$\text{V}_\text{g}$$\leq$$+30$ V} was used to modulate its charge carrier concentration. 

Magnetoresistance measurements were carried out at $\text{T} = 4.2$ K with pulsed magnetic fields up to $\text{B} = 55$ T. To avoid thermal stress, the sample was kept at constant temperature between measurements. The longitudinal magneto-resistance (MR) was positive at low magnetic fields and reached a broad maximum at $\text{B}_\text{max}$, which was followed by a region of negative slope. The main features of the measurements were reproducible during the increase and decrease of the magnetic field pulse. Hall resistance curves (presented in the suppl. Material \cite{suppl_material}) showed a pronounced non-linear behavior for all values of $\text{V}_\text{g}$, tending towards zero at large B. These results are qualitatively similar to observations in macroscopic graphite \cite{Kumar2010, Akiba2015}. Unfortunately, such behavior cannot allow for a reliable determination of the sample charge carrier concentration, as the simplest fitting of the Hall curves would rely on the two fluid model for which at least four independent parameters (carrier density and mobility for electrons and holes) are necessary \cite{Du2005}. The fitting procedure further gains in complexity in the realistic cases of magnetic-field-dependent mobility, occurrence of partial charge screening or when additional sub-band contributions are considered. %Qualitatively, however, the variation of $\text{R}_\text{xy}(\text{B})$ with $\text{V}_\text{g}$ allow us to infer the occurence of the modulation of charge carrier concentration in the device.

%In our sample, the HRS shows a critical field $\text{B}_\text{c}>38$ T. This corresponds to the magnetic field for which $\text{R}_\text{xx}(\text{B})$ deviates from its smooth background, as indicated in fig. \ref{fig_colormap}a. Such state is strongly influenced by the backgate voltage, as shown in fig. \ref{fig_colormap}. The results show a non-monotonic shift of $\text{B}_\text{c}$ and a change of the relative intensity of the HRS with $\text{V}_\text{g}$. In order to correlate such changes with the variation of the average charge carrier concentration in the material, we analyzed the Shubnikov de Haas (SdH) oscillations present in all measurements. Experiments revealed SdH oscillations with one dominant frequency component. The characteristic SdH frequencies ($\text{f}_\text{SDH}$) reached up to $75$ T when varying $\text{V}_\text{g}$ between $-30$ V and $+15$ V, as shown in fig. \ref{fig_analisys}a (the quantum oscillations are shown in the suppl. material \cite{suppl_material}). Such $\text{f}_\text{SDH}$ is tenfold higher than values typically found in the literature for pristine graphite (which ranges between $4.5$ T and $7$ T) \cite{Hubbard2011}. Yet, using the Onsagner relation ($\text{S}=2 \pi e \text{f}_\text{SDH}/\hbar$) we estimate the corresponding cross-section S of the Fermi surface (FS) at approx. $\text{S}\approx 7.15 \times 10^{-3} \text{ \AA}^{-2}$, which is still three orders of magnitude below the in-plane cross-section of the first Brillouin zone of graphite (of approx. $7.53 \text{ \AA}^{-2}$) \cite{Chung2002}.

\begin{figure}
	\includegraphics[width=8cm]{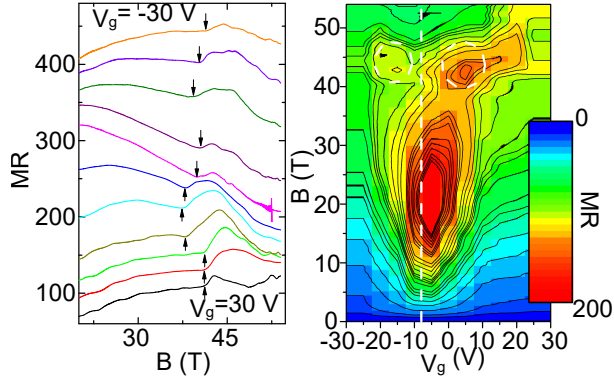}
	\caption{a) Magnetoresistance featuring the HRS, measured at $\text{T} = 4.2$ K at different backgate voltages (from top to bottom)  $\text{V}_\text{g} = -30$ V, $-20$ V, $-15$ V, $-10$ V, $-5$ V, $0$ V, $5$ V, $10$ V, $15$ V, $20$ V, $30$ V. The curves have been shifted vertically for clarity. The arrows show the approximate values of $\text{B}_\text{c}$ for each $\text{V}_\text{g}$. b) colormap of the magnetoresistance measurements (see the suppl. material for individual curves \cite{suppl_material}). The dashed line corresponds to the voltage necessary to put the sample in a compensated situation (see the text). Note the occurrence of more intense HRS to the left and to the right of the dashed line (marked by dashed circles).}
	\label{fig_colormap}
\end{figure}

\begin{figure}
	\includegraphics[width=8cm]{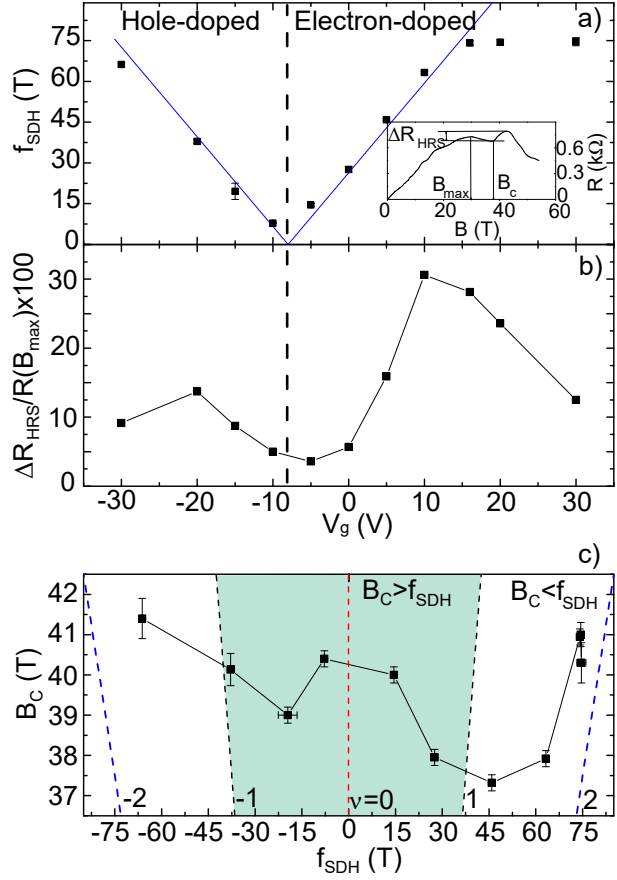}
	\caption{a) SdH frequency $\text{f}_\text{SDH}$ and b) relative amplitude of the HRS as a function of the backgate voltage $\text{V}_\text{g}$. The lines in a) are a function of the type $\text{f}_\text{SDH} = \beta |\text{V} - \text{V}_\text{CNL}|$, with $\text{V}_\text{CNL} = -7.6$ V and $\beta = 3.3$ T/V. The inset shows how $\Delta \text{R}_\text{HRS}$ and $\text{B}_\text{max}$ were defined. The relative HRS intensity was defined as $\Delta \text{R}_\text{HRS}/\text{R}(\text{B}_\text{max})$. c) Triggering magnetic field $\text{B}_\text{c}$ as a function of the SdH frequency. Negative frequencies denote SdH oscillations due to holes. The dashed lines correspond to the magnetic fields necessary to achieve the integer filling factors $\nu(\text{B})\equiv \text{f}_\text{SDH}/\text{B}=0$,$\pm 1$,$\pm 2$ , which correspond to the number of Landau levels occupied at certain B. Points outside the shaded area correspond to the HRS being triggered outside the quantum limit ($\text{B}_\text{c} < \text{F}_\text{SDH})$}
	\label{fig_analisys}
\end{figure}

In our sample, the HRS shows a critical field $\text{B}_\text{c}>38$ T. This corresponds to the magnetic field for which $\text{R}_\text{xx}(\text{B})$ deviates from its smooth background, as indicated in fig. \ref{fig_colormap}a. Such state is strongly influenced by the backgate voltage, showing a variation of its relative intensity and a non-monotonic shift of $\text{B}_\text{c}$ with $\text{V}_\text{g}$. In order to correlate such changes with the alteration of the charge carrier concentration in the material, we analyzed the Shubnikov de Haas (SdH) oscillations present in all measurements. To determine their frequencies, we employed the method used in refs. \cite{Zhang2005, Jobiliong2007}. Unfortunately, this method only yields the value of the dominant SdH frequency ($\text{f}_\text{SDH}$), specially if other components have a much lower intensity. The obtained $\text{f}_\text{SDH}$ reached up to $75$ T when varying $\text{V}_\text{g}$ between $-30$ V and $+15$ V, as shown in fig. \ref{fig_analisys}a (the quantum oscillations are shown in the suppl. material \cite{suppl_material}). Such $\text{f}_\text{SDH}$ is tenfold higher than values typically found in the literature for pristine graphite (which ranges between $4.5$ T and $7$ T) \cite{Hubbard2011}. We did not observe sharp variations of the oscillations's Berry phase with $\text{V}_\text{g}$ (see the suppl. material \cite{suppl_material}).

Our results show a linear dependency between $\text{f}_\text{SDH}$ and $\text{V}_\text{g}$ for \mbox{$\text{V}_\text{g}<15$ V}, characterizing the sample as a quasi-2D system \cite{Ashcroft}. For $\text{V}_\text{g}>+15$ V, $\text{f}_\text{SDH}$ became almost constant with $\text{V}_\text{g}$, presumably due to enhanced Coulomb screening caused by an excess of charge carriers. A linear extrapolation of the data for $\text{V}_\text{g}<15$ V found $\text{f}_{\text{SDH}}=3.3\left|\text{V}_\text{g}+7.6 \right|$ ($\text{f}_{\text{SDH}}$ given in Tesla and $\text{V}_\text{g}$ in volts), suggesting that compensation in our device (equivalent electron and hole population) occurs around $\text{V}_\text{CNL}= -7.6$ V. This is shown in fig. \ref{fig_analisys}a. At this voltage, $\text{B}_\text{c}$ shows a local maximum, presenting minima at the nearly symmetrical values with respect to $\text{V}_\text{CNL}$ $\text{V}_\text{g} \approx -15$ V and $\text{V}_\text{g} \approx +5$ V. At these values we also observed an increase of the relative intensity of the HRS ($\Delta \text{R}_\text{HRS}/\text{R}(\text{B}_\text{max})$, see fig. \ref{fig_analisys}b), followed by its reduction at larger doping. This observation is highlighted in the colormap of fig. \ref{fig_colormap}b, which shows two local maxima above $40$ T caused by the HRS (marked by dashed circles), quasi-symmetric with respect to the CNL. %If one considers that the HRS has origin in electron-electron interactions \cite{Yaguchi2009, Arnold2017}, the suppression of its amplitude at large $\text{V}_\text{g}$ corroborates the occurrence of the enhanced Coulomb screening as suggested by the results of figure \ref{fig_analisys}a.

We can account for the occurrence of charge screening in our device by associating the modulation of $\text{f}_\text{SDH}$ to a change of the tree-dimensional charge carrier density (electrons or holes) $\eta_{\text{3D}}$ in the sample. For this, we use a simplified model for a quasi-2D electron gas
\begin{equation}
\eta_{\text{3D}}=2e\frac{S}{(2\pi)^2}\frac{1}{c_0}\alpha.
\label{eq_charge_concentration}
\end{equation}
In it, $e$ is the electronic charge, $S$ is the 2-dimensional in-plane Fermi surface cross-section, $\alpha$ is a constant related to the dimensionality of the material and $c_0 = 0.335$ nm the interlayer distance in graphite. Under the assumption of decoupled graphene layers, $\alpha$ = 1, whereas if one considers the existence of a closed Fermi surface with highly anisotropic ellipsoidal electrons (or holes) pockets spanning half of the unit cell along $k_z$, $\alpha \approx 2/3$. Using the Onsagner relation and assuming in-plane isotropy, the cross-section S of the Fermi surface relates to $\text{f}_\text{SDH}$ according to $S = (2\pi)^2 e/h \times \text{f}_\text{SDH}$.

Multiplying both sides of eq. \ref{eq_charge_concentration} by the thickness $t$ of the sample and replacing $\eta_{\text{3D}}\times t \equiv \eta_{\text{2D}} = \text{C}_\text{SDH}|\text{V}_\text{g}-\text{V}_\text{CNL}|$, we obtain an expression relating $\text{V}_\text{g}$ to $\text{f}_\text{SDH}$ through a specific capacitance per unit of area $\text{C}_\text{SDH}$. The latter can be isolated, resulting in
\begin{equation}
\text{C}_\text{SDH}= 3.3 t \frac{2e^2\alpha}{hc_0},
\label{eq_charge_concentration2}
\end{equation}
where the numerical prefactor (3.3) is given in T/V and corresponds to the slope extracted from the linear fit presented in the $\text{f}_\text{SDH}$ vs. $\text{V}_\text{g}$ plot (fig. \ref{fig_analisys}a).

Using expression \ref{eq_charge_concentration2} with $t = 35$ nm and $\alpha = 2/3$, we obtain $\text{C}_\text{SDH} \approx 1.7 \times 10^3$ nF/$\text{cm}^2$. This value is two orders of magnitude above the specific capacitance estimated from geometrical considerations $\text{C}_\text{g} \approx 11$ nF/$\text{cm}^2$. Such result suggests that the region subject to gating is confined to a small portion of the graphite sample, close to its interface with the BN substrate. The effective thickness of this region can be estimated by finding $t=t_{eff}$ for which $\text{C}_\text{SDH} = \text{C}_\text{g}$ in eq. \ref{eq_charge_concentration2}. This method yields $t_{eff} \approx 0.2$ nm, which is below the currently accepted screening length for graphite (of $\lambda_s = 0.4$ nm) \cite{Zhang2005}. The small value obtained for $t_{eff}$ can be attributed to the approximated model employed here, as the parameter $\alpha$ in eq. \ref{eq_charge_concentration} is unknown.   %For simplicity, three-dimmensional charge carrier concentration in eq. /ref{eq_charge_concentration} can be reduced to an effective two-dimensional carrier concentration by performing $\eta_{\text{2D}}=\eta_{\text{3D}}\times t$, $t$ the thickness of the sample. 

Another estimation for $t_{eff}$ can be obtained from a screened, simple two-band model proposed for graphite in ref. \cite{Zhang2005}. Using our sample parameters and $\lambda_s = 0.4$ nm as the conventional screening length, we obtain a second estimation for $t_{eff}$ at $\text{V}_\text{g} = 15 V$ of, at most, $t_{eff}\approx 2.5 $ nm (independent of B). The calculated values correspond to the distance from the interface between graphite and BN above which the gating effect is \textit{completely} screened. Portions of graphite further from the BN interface are expected to be outside the influence of any electrostatic modulation. We briefly discuss the limitations of this model in the suppl. material.

Despite small estimated values of $t_{eff}$, the prominent occurrence of SdH oscillations and the reversal of the Hall signal with $\text{V}_\text{g}$ qualitatively suggest that the gated portion of the sample dominates the magnetotransport properties of the device. For the purposes of our discussion, we consider such region to have a thickness of, at most, $2.5$ nm (the largest of our previous estimations). Even though confined, carriers in this region must present an off-plane degree of freedom, as no signatures of a quantum-Hall state typical of few-layer graphene are observed in our measurements\cite{Novoselov2006, Kumar2011}.

We model transport in our system as two contributions occurring in parallel: one from the doped region ($\text{G}_1$) and the other from the screened portion graphite ($\text{G}_2$). The measured conductivity of the sample can, then, be described as
\begin{equation}
\text{G}_\text{tot}=\text{G}_1+\text{G}_2.
\label{eq_Gtot}
\end{equation}
As the screened region of the sample ($\text{G}_2$) is not expected to be affected by $\text{V}_{\text{g}}$, our results can be understood as a consequence of gating over a small fraction of the sample, measured in parallel to a background signal of pristine graphite. The modulation of the HRS and its earlier triggering at different $\text{V}_\text{g}$, therefore, can be attributed to variations of $\text{G}_1$ in eq. \ref{eq_Gtot}.

Under these assumptions, the thin gated region in the sample does not support an exclusive off-plane origin for the HRS. Within the context of a c-axis CDW transition, the typical off-plane lengths associated with the phenomenon are expected to be around $1.8 \text{ nm} \leq \lambda_{\text{CDW}} \leq 2.28 \text{ nm}$, depending on the charge carrier considered (electrons or holes) \cite{Yoshioka1981, Arnold2017, Jobiliong2007}. The clear modulation of the HRS in a region of similar thickness (max. $2.5$ nm) suggests a state possessing also a strong in-plane component. This hypothesis is supported by measurements in a $4$ nm thick device (presented later on), which did not show signs of the HRS. If such thickness represents a minimum c-axis dimension necessary to trigger the phenomenon, one would not expect the variation of the charge carrier concentration in a region of similar dimension to modulate the HRS.

It can be argued that another possibility to account for our observations is to consider an earlier triggering of the HRS due to the presence of a charge gradient between the screened electron/hole gas and graphite. For high (low) enough gate voltages, the  region of the sample nearest to the interface with BN is expected to be completely depleted of electrons (holes). As the distance from the gate is increased, however, the Fermi level shifts and both graphite's electronic bands become populated, eventually reaching a region unaffected by gating \cite{Zhang2005}. Under this hypothesis, however, the HRS cannot be described as an off-plane effect. Within the context of a 1D c-axis transition, the the conditions necessary for the earlier triggering would happen locally (on an infinitesimally thin region) - resulting in a contradiction. In this case, the HRS could be described exclusively as an in-plane effect.

We also note that the achievement of the quantum limit was not a requirement for the HRS in our device. This is illustrated in fig. \ref{fig_analisys}c. The data shows that the HRS takes place at a ``filling factor'' (which describes the number of occupied Landau levels) $\nu \equiv \text{f}_\text{SDH}/\text{B} > 1$, or alternatively $\text{B}_\text{c} < \text{f}_\text{SDH}$. This observation suggests that multiple Landau levels (both electron-like and hole-like with $n \neq 0$) could host the conditions necessary for triggering the phenomenon. These results are at odds with current models employed for the HRS in graphite, which predict that the effect takes place as a consequence of instabilities in the Fermi surface, expected only when the lowest LL is populated \cite{Yaguchi2009, Akiba2015}.

Despite our results suggesting that the HRS has a strong in-plane nature and that the quantum limit is not a requirement for it, we observe that the earlier triggerings of the HRS happen nearly symmetrically with respect to the CNL (see fig. \ref{fig_analisys}c). In our understanding, such bipolar behavior demonstrates that the HRS is triggered by ideal concentrations of both electrons and holes in the material.  Assuming an origin due to selective population/depopulation of LLs, the non-monotonic response of $\text{B}_\text{c}$ to $\text{V}_\text{g}$ can be understood in terms of a selective tuning of the occupation of each LL. This causes the conditions for the HRS to be achieved at different magnetic fields when varying $\text{V}_\text{g}$. In this aspect, our results agree with the hypothesis that the HRS involves both electron and hole subbands in graphite, as proposed by Fauque et al. in \cite{Fauque2013}.

Using a simple model for a 2D electron gas, and assuming the in-plane effective masses of electrons and holes in graphite at $m^* = 0.07$ $m_e$ and $0.04$ $m_e$ respectively \cite{Arnold2017} ($m_e$ the electron rest mass), we further estimate the Fermi energies $\text{E}_\text{F}$ for each $\text{f}_\text{SDH}$:
\begin{equation}
	\text{E}_\text{F} \approx \frac{e\hbar}{m^*} \text{f}_\text{SDH}.
	\label{eq_FE}
\end{equation}         
Using this expression, we obtain that that the triggering of the HRS happen nearly symmetrical around the CNL, at $\text{E}_\text{F} \approx ~ \pm 500$ meV. These energy values are in strong disagreement with band calculations at high magnetic fields assuming a c-axis CDW scenario at the lowest Landau level, for which tenths of meV are expected \cite{Yaguchi2009, Arnold2017, Sugihara1984}. We attribute such discrepancy to the overly simplified model employed here (which captures only qualitatively the quasi-2D character of graphite), as well as to the apparent strong in-plane component of the HRS which is unaccounted for in the description of the phenomenon.

Interestingly, our results do not support the hypothesis of a magnetic-field-induced excitonic gap as the cause for the HRS in graphite, as recently suggested in reference \cite{Zhu2017}. In this scenario, the symmetric triggering of the phenomenon with respect to the CNL - as well as the non-monotonic dependency of its relative amplitude - would require graphite to present a complex c-axis band structure, featuring multiple gaps (not predicted in ref. \cite{Zhu2017}).

We do not discard, however, that both an excitonic gap or a CDW state could take place in graphite prior to the quantum limit. In this case, multiple Landau levels would be involved, adding much complexity to the description of the phenomenon. Indeed, systems presenting CDW states at filling factors above $1$ are expected to show exotic properties, such as the occurrence of bubble phases (Wigner crystallization at high LLs) in quantum Hall gases \cite{Eisenstein2000, Koulakov1996, Zhang2007}. Such phenomenon is predicted to take place in graphene at fractional filling factors above the n=1 LL \cite{Zhang2007}. Theoretically, a direct evidence of such state could be detected by verifying the in-plane anisotropy of the sample resistivity during the HRS at $\nu > 1$ \cite{Zhang2007, Eisenstein2000}. In this case, the HRS could be decisively identified as having a purely in-plane origin, rather than c-axis. Unfortunately, the geometry of our device did not allow for such verification. 

%A simmilar phenomenon is predicted to take place in ABC graphene   predicted in graphene in the quantum Hall regime.   Among the possibilities justifying a CDW in this context include the creation of a Wigner crystal in the material, in which case the band structure of the material and allowing for additional nesting vectors. In this context, 

Our experiments highlight another relevant aspect of the HRS in HOPG. Observations scattered across the literature reveal a wide range of triggering magnetic fields $\text{B}_\text{c}$ (as low as $22$ T and as high as $45$ T) \cite{Timp1983, Nakamura1983, Kumar2010, Fauque2013,  Yaguchi2009, Yoshioka1981, Akiba2015, Arnold2017}. While these values are usually deemed sample dependent, our results strongly suggest that the intrinsic charge carrier density in graphite plays a major role in its determination, as well in the amplitude of the HRS. This becomes important when comparing in- and out-of-plane measurements to obtain the direction in which the HRS takes place. Unless performed in the same physical sample, these comparisons must carefully account for the transport properties of each crystal individually, which can be easily overlooked in c-axis measurements due to the lack of an $\text{R}_{\text{xy}}$ component. Greater care is necessary if different values of $\text{B}_{\text{c}}$ are obtained in different experiments.

Indeed, conflicting estimations in the literature provide a 2D, low temperature charge carrier density $n_{2D}$ in graphite ranging from $10^{10}$ to $10^{12}$ $\text{cm}^{-2}$ \cite{Garcia2008, Klein1961, Arndt2009}. In the experiments shown here, $\text{V}_\text{g}$ causes an estimated charge carrier variation of one order of magnitude (between $n_{2D}\approx 10^{11}$ and $10^{12}$ $\text{cm}^{-2}$, estimated from $n\approx\text{f}_\text{SDH}\times2e/h$), resulting in a shift of $\text{B}_\text{c}$ between $37$ T and $41$ T. We conjecture that a further increase of the charge carrier concentration in our sample would cause the HRS to be triggered for even larger values of $\text{B}_\text{c}$. Certainly, disorder is also expected to play an important role, as it causes an effective broadening of Landau and sublandau levels in the material, smearing out measurable quantum phenomena \cite{Yaguchi2009}. However, our experiments were not designed to verify such impact.

We close our work by addressing the influence of the sample dimensionality in the HRS of graphite. It is widely accepted that the HRS is an off-plane phenomenon \cite{Kumar2010, Fauque2013, Yaguchi2009, Yoshioka1981, Sugihara1984, Taen2018}, with different reports in the literature demonstrating a larger relative intensity along c-axis measurements (see, for example, \cite{Fauque2013} and references therein) - albeit in different samples. Its presence in in-plane measurements is usually explained in terms of the 3D Fermi surface of graphite, in which a gap opening along the sample’s c-axis affects the in-plane dispersion as well.

As previously discussed, however, our results suggest that the modulation of the HRS takes place in a graphite region with thickness inferior to $2.5$ nm. Such dimension is less than two times the characteristic wavelengths expected for a c-axis transition, suggesting that variations observed as a function of $\text{V}_\text{g}$ happen due to the phenomenon possessing a strong in-plane character. Such picture was proposed during early experimental works on the subject \cite{Timp1983}, although, to our best knowledge, current models view the HRS as a phenomenon triggered in the c-axis direction \cite{Kumar2010, Fauque2013, Akiba2015, Arnold2017, Takahashi1994, Taen2018}.  

\begin{figure}
	\includegraphics[width=8cm]{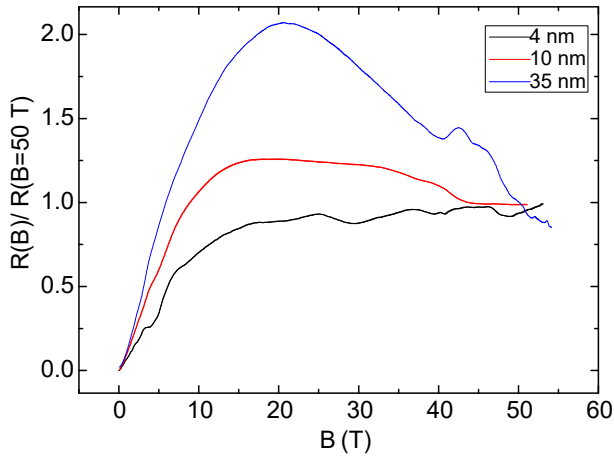}
	\caption{Magnetoresistance curves normalized at $\text{R}(\text{B} = 50 \text{ T})$ for samples of different thicknesses. Note the suppression of the HRS upon thinning down the sample. Samples $35$ nm and $10$ nm were measured at liquid He temperature ($\text{T} = 4.2$ K), whereas the sample with $4$ nm thickness was measured at $\text{T} = 300$ mK.}
	\label{fig_thickness}
\end{figure}

In an attempt to probe an eventual in-plane character of the HRS, magnetoresistance measurements were performed in additional HOPG samples with thicknesses $t = 10$ nm and $t = 4$ nm. These correspond to devices with, approximately, $12$ and $30$ graphene planes, respectively. All the samples were exfoliated from the same bulk crystal and measured when undoped. They presented $\text{f}_\text{SDH} = 5.3 \pm 0.1$ T and $7.14 \pm 0.05$ T ($t=4$, $10$ nm, respectively), thus suggesting charge carrier concentrations of the same magnitude as observed in the \mbox{$t=35$ nm} device at $\text{V}_\text{g} = -10$ V (Near the CNL - see the suppl. material for the quantum oscillations \cite{suppl_material}). Unfortunately, we have no control over $\text{f}_\text{SDH} (\text{V}_\text{g}=0)$ at samples with different thickness, which could be caused by different factors such as interaction with the substrate, charged impurities, capacitive effects or be inherent of the graphite used. Results are shown in fig. \ref{fig_thickness}. 

A progressive suppression of the HRS is observed in our devices, with the $10$ nm sample presenting a strongly attenuated feature around $\text{B} \approx 37$ T and the $4$ nm one not showing any signatures of the phenomenon. We argue that the suppression of the HRS in our 4 nm sample cannot be attributed to enhanced structural disorder, as the device has tell-tale indicatives of high quality such as clear quantum oscillations and signatures of a fractional quantum Hall state (see the suppl. material \cite{suppl_material}). These observations are at odds with previous measurements reporting the absence of the HRS on mesoscopic graphite thinner than $130$ nm \cite{Jobiliong2007}. Such discrepancy can be attributed to the magnetic field range employed in ref. \cite{Jobiliong2007} (below 35 T), which probably was below the limit necessary to trigger the HRS in their devices.

We also compare our results to a recent study by T. Taen et al. \cite{Taen2018}, in which the effect of sample thickness is considered. In their work, the HRS is modeled as a c-axis transition and its suppression in thinner devices is attributed to the quantization of the  $k_z$ wave vector due to confinement. Experimentally, this should manifest as a displacement of $\text{B}_\text{c}$ towards larger values of B as devices are made thinner. 

Our results indeed suggest that finite-size effects play a role in suppressing the HRS, as the lack of a HRS in our $t=4$ nm device can be attributed to confinement effects settling in, which effectively reduce the dimensionality of the material by inhibiting off-plane transport. However, our experimental data does not behave qualitatively as predicted by the model presented in ref. \cite{Taen2018}. In particular, there is no increase of $\text{B}_\text{c}$ in the $t=10$ nm device in comparison to the $t=35$ nm device - i.e. $\text{B}_\text{c}(t=35 \text{ nm}) \geq \text{B}_\text{c}(t=10 \text{ nm})$, with $\text{B}_\text{c}(t=35 \text{ nm}) \approx \text{B}_\text{c}(t=10 \text{ nm}) \approx 37.5$ T achieved for $\text{V}_\text{g}= 5$ V (far from the CNL). In addition, the $\text{B}_\text{c}(T)$ behavior on the $t=35$ nm sample does not seem to be compatible with the phase diagram presented in ref. \cite{Taen2018} (partially reproduced in the suppl. material \cite{suppl_material}), although this comparison might be inappropriate due to the different types of graphite used.

Instead, the absence of the phenomenon in the $4$ nm thick sample and its modulation in a region below $2.5$ nm  in our $t = 35$ nm device strongly suggest that an off-plane degree of freedom is essential for the HRS to take place, even if a strong in-plane contribution is present. This can be understood in the context of quantum fluctuations present in 2D systems, which can frustrate long-range interactions. A progressive addition of a third dimension, in this case, quenches (stabilizes) the strong two-dimensional quantum fluctuations \cite{Hohenberg1967}, allowing in-plane correlations leading to the HRS to take place.

\section{Conclusions}
In conclusion, in this work, we studied the HRS in graphite by modulating the charge carrier concentration of a mesoscopic sample. To our best knowledge, this report constitutes the first experimental observation of the influence of electrostatic doping on the HRS, which was triggered symmetrically with respect to the CNL in the sample. The amplitude and critical magnetic fields $\text{B}_\text{c}$ associated with it were strongly affected by the sample charge carrier concentration, which partially explains conflicting reports in the literature.  Taking into consideration the small fraction of the sample affected by the electrostatic doping, our results strongly suggest that the HRS has a large in-plane contribution. However, measurements in samples with different thicknesses demonstrate that an off-plane degree of freedom is essential to stabilize the phenomenon, whose presence is reported in samples as thin as $10$ nm. Our experiments also allowed us to infer that the achievement of the quantum limit is not a necessary condition for the occurrence of the HRS, opening the possibility that multiple Landau Levels (with $n \geq 1$) might be involved. These observations are at odds with the current understanding of the phenomenon, in which the HRS is associated to a c-axis electronic phase transition occurring above the quantum limit. We expect our work to inspire similar investigations, which could further clarify the properties of the HRS for $\nu>1$.% Assuming that the HRS in graphite has origin due to a CDW transition, we experimentally  estimate an upper and lower and limit for a c-axis nesting wavevector in the material at $0.063 \text{ \AA}^{-1}<2k_\text{F}< 0.312 A^{-1}$. 

\section*{Acknowledgments}
We would like to thank B. Fauque and Y. Kopelevich for fruitful discussions. This work has received funding from the People Programme (MarieCurie Actions) of the European Union's Seventh Framework Programme (FP7/2007-2013)under REA grant agreement n. PCOFUND-GA-2013-609102, through the PRESTIGE programme coordinated by Campus France. This study has been partially supported through the grant NEXT n° ANR-10-LABX-0037 in the framework of the ``Programme des Investissements d'Avenir''. B. C. C. acknowledges financial support from the Polish National Science Center under project number UMO-2014/15/B/ST3/03889. High magnetic field measurements were performed at LMCMI under proposal TMS02-215.
%\end{acknowledgments}

\end{document}